\documentclass[aps,pra,twocolumn,superscriptaddress]{revtex4-2}
\usepackage{bm}
\usepackage{graphicx}
\usepackage{amssymb,amsmath,amsbsy,amsgen,amsfonts}
\usepackage{dcolumn}
\usepackage{amsthm}
\usepackage{mathrsfs}
\usepackage{latexsym}
\usepackage{array}
\usepackage{amstext}
\usepackage{epsfig}
\usepackage{epstopdf} 

\usepackage{color}
\usepackage{units}
\usepackage{calrsfs}

\newcommand{\be}{\begin{equation}}
\newcommand{\ee}{\end{equation}}
\newcommand{\ba}{\begin{array}}
\newcommand{\ea}{\end{array}}
\newcommand{\bqa}{\begin{eqnarray}}
\newcommand{\eqa}{\end{eqnarray}}

\begin{document}

\title{Dynamical Casimir Effects: \\The Need for Nonlocality in Time-Varying Dispersive Nanophotonics}

\author{S. Ali Hassani Gangaraj} \email{ali.gangaraj@gmail.com}
\address{Optical Physics Division, Corning Research and Development, Sullivan Park, Corning, New York 14831, USA}

\author{George W. Hanson} \email{george@uwm.edu}
\address{Department of Electrical Engineering, University of Wisconsin-Milwaukee, Milwaukee, WI 53211, USA}

\author{Francesco Monticone} \email{francesco.monticone@cornell.edu}
\address{School of Electrical and Computer Engineering, Cornell University, Ithaca, NY 14853, USA}

\date{\today}

\begin{abstract}

Both real and virtual photons can be involved in light-matter interactions. A famous example of the observable implications of virtual photons --- vacuum fluctuations of the quantum electromagnetic field --- is the Casimir effect. Since quantum vacuum effects are weak, various mechanisms have been proposed to enhance and engineer them, ranging from static, e.g., strong optical resonances, to dynamic, e.g., systems with moving boundaries or time-varying optical properties, or a combination of them. In this Letter, we discuss the role of material nonlocality (spatial dispersion) in dynamical Casimir effects in time-varying frequency-dispersive nanophotonic systems. We first show that local models may lead to nonphysical predictions, such as diverging emission rates of entangled polariton pairs. We then theoretically demonstrate that nonlocality regularizes this behavior by correcting the asymptotic response of the system for large wavevectors and reveals physical effects missed by local models, including a significant broadening of the emission rate distribution, which are relevant for future experimental observations. Our work sheds light on the importance of nonlocal effects in this new frontier of nanophotonics.

\end{abstract}

\maketitle


The energy levels of a quantum mechanical resonator are given by $E_n = (n+1/2) \hbar \omega $, which implies a non-vanishing minimum energy (the zero-point energy) in the ground state $ n = 0 $ \cite{Wallas}. Indeed, if the Hamiltonian of a harmonic oscillator, $ H = p^2/2m + m\omega^2z^2/2 $, were zero, then both position $z$ and momentum $p$ would need to be identically zero, which would violate Heisenberg's uncertainty principle. Hence, both $z$ and $p$ must \emph{fluctuate} in the ground state, producing \emph{zero-point fluctuations}. Each mode of the electromagnetic field is analogous to a quantum harmonic oscillator, thus the quantized field exhibits zero-point fluctuations (virtual photons) in the lowest energy state, i.e., the \emph{vacuum} state. Since the quantum electromagnetic vacuum fluctuates, it interacts with atomic systems even in the absence of real photons \cite{Welsch}, thereby giving rise to many observable effects including van der Waals and Casimir forces \cite{Lamoreaux}, spontaneous emission \cite{Weisskopf,Purcell,Gerard}, the Lamb shift \cite{Lamb}, non-contact quantum friction \cite{Pendry,Kardar}, and all the way up to its potential role in the evaporation of black holes through Hawking radiation \cite{Lamoreaux,ModernPhys} and the cosmological constant problem \cite{Weinberg}.

Among all these phenomena, the most well-known is arguably the Casimir effect, described by Hendrik Casimir in the 1940s \cite{Casimir_1, Casimir_2, Casimir_3}. Since the strength of the Casimir (and related) effects is determined by how the electromagnetic modes are modified and constrained by the considered structure, they can be controlled by engineering the geometry, material, and symmetries of the system. For example, it has been shown \cite{Purcell,Rodriguez,Volokitin} that strong photonic resonances in nanophotonic structures and polaritonic materials, as well as breaking reciprocity through magneto-optic effects \cite{Hassani_Fluctuation,Hassani_PRA,Hassani_PRB,Hassani_Spontaneous}, can be used to enhance and engineer Casimir forces/torques and other vacuum fluctuation effects. In addition to these  ``static'', i.e., time-invariant, mechanisms, also ``dynamical'' approaches, based on moving media or time-modulated systems, have been investigated to amplify fluctuation-based phenomena and convert virtual photons into entangled pairs of real photons \cite{ModernPhys}. Examples include cavities with oscillating boundaries \cite{Moore}, spatially localized index perturbations accelerated on some trajectory \cite{accelaration}, and time-modulated polar insulator thin-films \cite{Jamison}. 

Investigating the effects resulting from vacuum fluctuations in dynamically modulated structures becomes especially complicated when the material optical properties are simultaneously temporally dispersive and time-dependent \cite{Jamison,Koutserimpas}. Particularly interesting and subtle are scenarios in which the time-modulation frequency becomes comparable with the material optical transition frequencies, where an ``adiabatic'' description of the time-varying material cannot be used. In Ref. \cite{Jamison}, an elegant theoretical framework was proposed to study dynamical Casimir effects in these scenarios, showing that pairs of entangled phonon-polaritons are generated on thin films of polar insulators when the transverse optical (TO) phonon frequency is rapidly modulated in time. 

In this Letter, we revisit this pair generation process and also investigate the decay rate of a two-level system in a time-modulated photonic environment that is not only temporally dispersive, but also \emph{spatially} dispersive, i.e., its spatial impulse response is \emph{nonlocal} \cite{Raza2015,Polar-Dielectrics,Monticone2020,Shastri2023,Khurgin}. We demonstrate that taking into account the natural nonlocal response of the material is essential in accurately describing dynamical Casimir effects in these systems. Ignoring nonlocality may lead to nonphysical behaviors.

\begin{figure}[h!]
	\noindent \includegraphics[width=1.0\columnwidth]{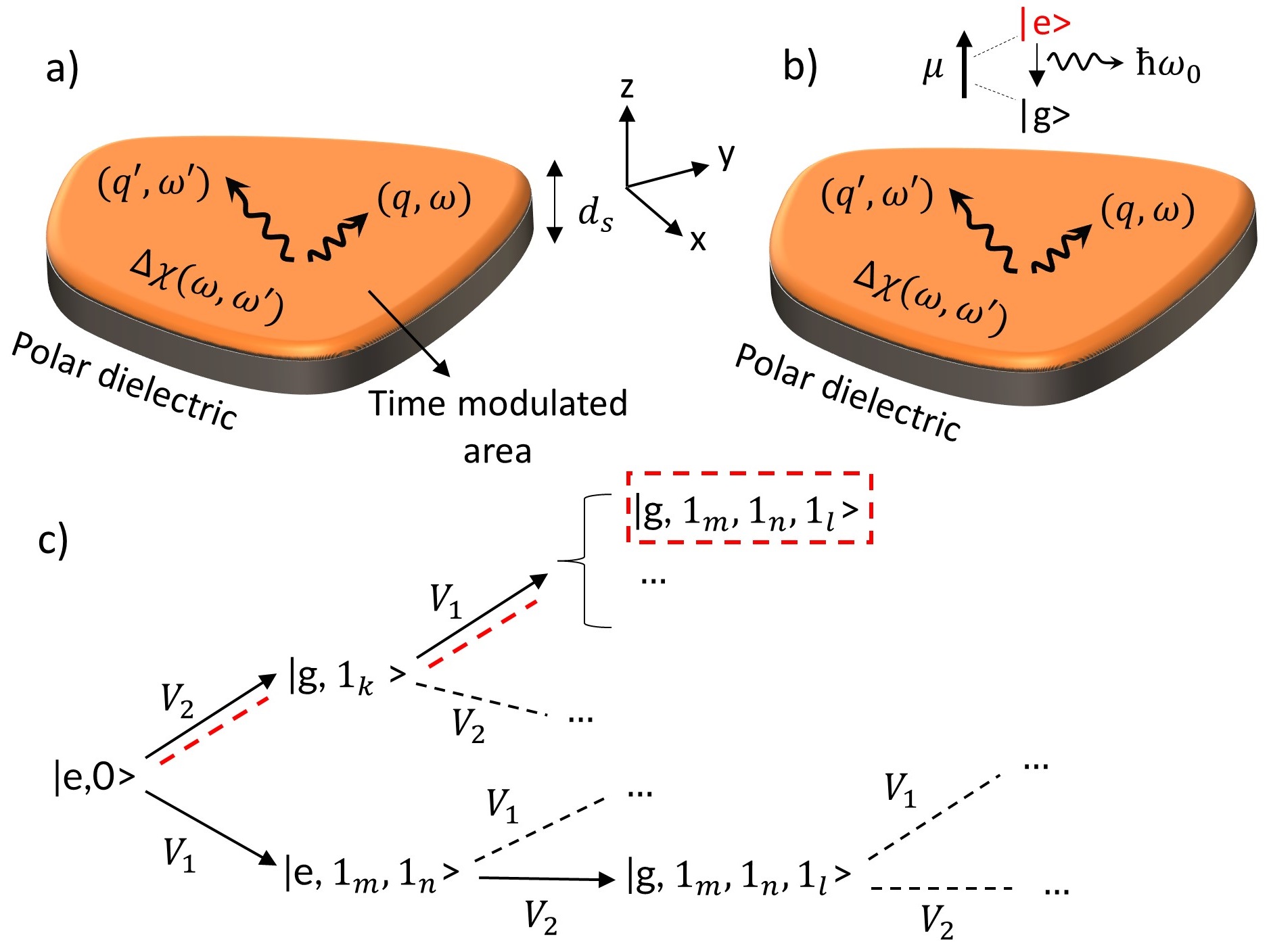}
	\caption{(a) A polar dielectric slab subject to a time modulation of its permittivity over a thin layer. The polar dielectric is supposed to be silicon carbide (SiC). (b) Time-modulated slab interacting with a two-level dipole source. (c) Hilbert space of the configuration in panel (b).}
	\label{fig1p}
\end{figure}

The geometry under consideration is shown in Fig. \ref{fig1p}(a): an isotropic slab made of a polar dielectric, with overall thickness $d_s$, is temporally modulated over a thin layer of thickness $d$. The time-invariant permittivity of the material, in the spatially local case, follows a standard Lorentz oscillator model, $ \epsilon_{bg}(\omega) = \epsilon_{\infty} \left( 1 + \omega_p^2 / \left(  \omega_0^2 - \omega^2 - i\gamma \omega \right)  \right)$, where $ \epsilon_{\infty} $ is the permittivity at high frequencies, $ \omega_0 $ is the  TO phonon frequency, $ \omega_p $ is the plasma frequency (oscillator strength), and $ \gamma $ is the damping rate. When a (weak) temporal modulation is applied to this dispersive material, the frequency-domain perturbed permittivity can be written as \cite{Jamison}
\begin{equation} \label{permittivity}
\epsilon(\omega , \omega') = \epsilon_{bg}(\omega)   2\pi \delta(\omega - \omega')  + \Delta \chi (\omega , \omega').
\end{equation}
Note the two-frequency formulation commonly used for systems with a time-varying linear response function in the non-adiabatic regime \cite{Koutserimpas}. In a time-varying dispersive planar structure as in Fig. \ref{fig1p}, the probability of pair generation, per unit frequency, per unit area of the sample, was then shown in Ref. \cite{Jamison} to be given by the following expression (see also \cite{SM})
\begin{align}\label{P}
	\frac{1}{A}  \frac{d P}{ d \omega} = & \int_{\omega' = 0}^{\omega' = \infty} d\omega'  \frac{ \left | \Delta \chi(\omega, -\omega')   \right|^2  }{16 \pi^3}   \int_{q = 0}^{ q = \infty}  d q ~ q \left( 1 - e^{-2qd}  \right)^2 \nonumber \\&   \mathrm{Im}(R_p( \omega', q )) \mathrm{Im}(R_p( \omega, q )) 
\end{align}
where $q$ is the wavenumber and $ R_p $ is the p-polarized reflection coefficient of the slab \cite{SM,Novotny,Two-photon}.

Inspecting Eq. (\ref{P}) reveals that this quantity diverges if the upper limit of the integral over wavenumber approaches infinity. Indeed, for the considered material model, $\mathrm{Im}(R_p)$ takes a finite nonzero value as $ q \rightarrow \infty $, whereas the wavenumber-dependent factor $q \left( 1 - e^{-2qd}  \right)^2$ never converges. A related issue is that, if arbitrarily large wavevenumbers need to be considered in Eq. (\ref{P}), $q$ would eventually become comparable to $ 2\pi/a_0 $, where $ a_0 \approx 10^{-11} ~\mathrm{m}   $ is the Bohr radius \cite{Wallas}. In this regime, this entire formulation would become inapplicable, as the validity of the dipole approximation itself would break down since the propagator factor, $ e^{iqx} $, would not be approximately constant over atomic/molecular scales (hence, multipolar responses would need to be included). 

We found that these issues do not appear only in the configuration considered above, but also in other related scenarios, such as in the problem of spontaneous emission from a two-level system in a time-modulated dispersive photonic reservoir. Considering a two-level atom radiating above a time-modulated polar dielectric slab, as in Fig. \ref{fig1p}(b), the total interaction Hamiltonian takes the form

\begin{align}
	& H_i(t)  = V_1(t) + V_2(t) = 	 - \epsilon_0 \int d\boldsymbol{r}  \int_{t'=-\infty}^{+\infty} dt' \Delta \chi_{ij}( \boldsymbol{r},t,t' ) \nonumber \\ &
\times E_j( \boldsymbol{r}, t' ) E_i ( \boldsymbol{r}, t )  - \boldsymbol{\mu} \cdot \boldsymbol{E},
\end{align} 
where $ \boldsymbol{\mu} $ is the dipole transition matrix of the two-level atom source. The sub-Hamiltonian $ V_1 $ consists of the operator terms $ a_ma_n, ~a_m^{\dagger}a_n, ~ a_m a_n^{\dagger},~ a_m^{\dagger} a_n^{\dagger} $, and $ V_2 $ contains terms like $ \sigma a_m^{\dagger} , ~ \sigma^{\dagger}a_m $, where $ \sigma,\sigma^{\dagger} $ and $ a,a^{\dagger} $ are the lowering/raising dipole and bosonic field operators, respectively, and $m,~n$ are mode indices. Assuming the system is initially prepared in the $ \left | e, 0 \right > $ state (excited atom and no photon in the bosonic mode), the total Hilbert space can be found by applying the interaction Hamiltonian to the initial state. As illustrated in Fig. \ref{fig1p}(c), an atom can make a transition from $  \left | e, 0 \right > $ to different states: Among them, it can be shown that the transition to $ \left | g, 1_m, 1_n, 1_l \right > $ (atom in ground state and photons at modes $m,n,l$) via the red path in Fig. \ref{fig1p}(c), occurs with the following decay rate (see \cite{SM}),
\begin{align}\label{gamma}
&	\gamma_{\left |e,0 \right> \rightarrow \left |g,1_m 1_n 1_l \right> } =  \frac{\gamma_0}{16 \pi ^3} \int d \omega'  \left | \Delta \chi(\omega_a, -\omega')   \right|^2    \nonumber \\& \int_{q=0}^{+\infty}  d q ~ q \left( 1 - e^{-2qd}  \right)^2  \mathrm{Im}(R_p( \omega', q )) \mathrm{Im}(R_p( \omega_a, q )),   
\end{align}
where $ \gamma_0 $ is the decay rate of an atom in the time-invariant photonic environment \cite{SM} and $\omega_a$ is the atomic transition frequency. By inspecting Eq. (\ref{gamma}), we note that similar issues exist as in the previous case, namely, the $q$-integral diverges and, at large enough wavenumbers, the dipole approximation validity would break down.

These issues are rooted in the dispersive nature of the considered time-varying material properties and how the asymptotic behavior for large wavenumbers is modeled. A local, wavevector-independent material model implies that the material polarization response persists for arbitrarily large wavevectors and predicts flat dispersion asymptotes for polaritonic modes such as surface plasmon- or phonon- polaritons, leading to unphysical behaviors and thermodynamic paradoxes even in the time-invariant case \cite{Fan,Optica,Monticone2020}.
A qualitative way to address this issue is to consider a local material response but set a large-wavevector cut-off in the form of a finite upper limit $q_c$ for the $q-$integral in Eqs. (\ref{P}) and (\ref{gamma}), as it has been done so far in the literature (e.g., \cite{Jamison}). While this approach is inspired by how physical nonlocal effects make the material response vanish for large wavevectors, difficulties may arise because (i) the choice of an upper limit can be rather arbitrary and, therefore, different values of $q_c$ would lead to different results, especially around material resonance frequencies, as shown in the following, and (ii) setting a sharp upper limit implies that the material polarization abruptly vanishes, which is not physical. Thus, assuming a local material model, with or without a sharp wavevector cut-off, may lead to potentially inaccurate results.  In the following, we show that these issues are more satisfactorily addressed by formally considering the impact of nonlocal effects on the response of dispersive time-varying materials.

\begin{figure}[h!]
	\noindent \includegraphics[width=1.0\columnwidth]{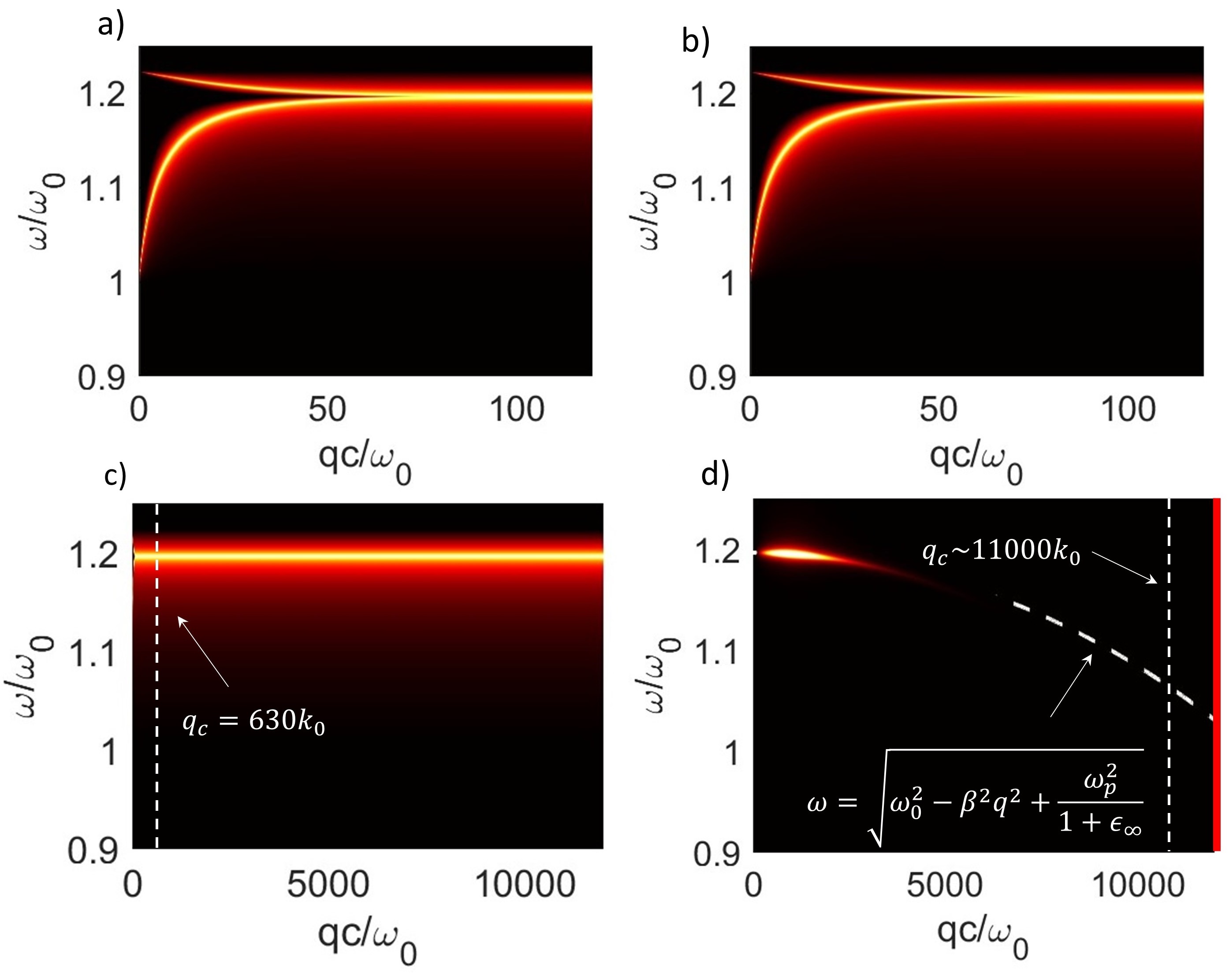}
	\caption{(a) Local ($ \beta = 0) $ and (b) nonlocal ($ \beta = 15.39\times 10^5~\mathrm{cm/s} $  \cite{Polar-Dielectrics}) dispersion diagrams for surface phonon-polaritons (SPhP) on a SiC slab with thickness $ d_s = 100 ~ \mathrm{nm}$. SiC dielectric parameters are taken from \cite{SiC}: $\omega_p = 1.049 \times 10^{14}    $ rad/s, $\omega_0 = 1.49 \times 10^{14}  $ rad/s, $ \epsilon_{\infty} = 6.7 $ and $ \Gamma = 8.97 \times 10^{11}  $ rad/s. (c), (d) Dispersion diagrams over a larger wavenumber range for the local and nonlocal cases, respectively. The dispersion diagrams are plotted as density plots of the magnitude of the slab reflectivity, with bright lines corresponding to SPhP poles (brighter colors mean higher intensity). The vertical dashed line in (c) marks the location of the wavenumber cut-off used for the local case in Fig. \ref{fig3p}(a). The vertical dashed line in (d) marks where the relevant integrals in the nonlocal case converge, whereas the solid red line marks where the wavenumber starts approaching the atomic scale ($ q \approx 2\pi/a_0 $).
	}
	\label{fig2p}
\end{figure}

A time-modulated polar dielectric at frequencies near its phonon-polariton resonance can be modeled as a lossy Lorentz oscillator with a time-varying resonant frequency: $ \omega_0 \rightarrow \omega_0 (1 + \delta \omega f(t)) $, where $ f(t) $ is a time-dependent function and $ \delta \omega \ll 1 $ is a small dimensionless parameter \cite{Jamison}. The equation of motion underlying this model is derived from Newton's Second Law for ions subject to a harmonic electric field, a time-varying restoring force, and a damping force. Here, to include nonlocality in a phenomenological manner \cite{Polar-Dielectrics}, a pressure term is also added, analogous to a hydrodynamic treatment of nonlocality for plasmonic materials \cite{Raza2015,Monticone2020,Shastri2023}, but with some important differences as discussed below and in \cite{Polar-Dielectrics,SM}. Thus, we obtain an equation for a nonlocal and time-varying Lorentz model,
\begin{align}
	\left( \frac{\partial^2 }{ \partial t^2 } + \gamma \frac{ \partial  }{ \partial t } + \omega_0^2 ( 1 + \delta\omega f(t))  + \beta^2 \nabla(\nabla \cdot )  \right) \frac{\textbf{p}(t)}{\epsilon_0\omega_p^2}  = \textbf{E}(t)
\end{align}
where $\textbf{p}(t)$ is the induced dipole moment and $\beta$ is a phenomenological velocity indicating the strength of nonlocality. For simplicity, we assumed the high-frequency nonresonant polarization response is negligible, i.e., $ \epsilon_{\infty} = 1 $. The above equation is the generalization of equation A1 in Ref. \cite{Polar-Dielectrics} for a polar dielectric material that is both nonlocal and, here, temporally modulated with a time-dependent TO resonant term. Assuming $ | \delta \omega | \ll 1 $, we show in \cite{SM} that the resulting spatially- and frequency-dispersive two-frequency permittivity can be written as in Eq. (\ref{permittivity}) with
\begin{align}\label{nonlocal-eps}
	&\epsilon_{bg}(\omega,q) = \left(1 + \frac{ \omega_p^2  }{(\omega^2_0(q) - \omega^2 - i\gamma \omega  }\right) \nonumber \\ & 
	\Delta \chi(\omega, \omega',q) =  \frac{  -\omega_0^2 \omega_p^2 \delta\omega f(\omega' - \omega) }{ (\omega^2_0(q) - \omega'^2 - i\gamma \omega' ) ( \omega^2_0(q) - \omega^2 - i\gamma \omega )}
\end{align}

where $ \omega^2_0(q) = \omega_0^2 - \beta^2q^2 $, and $ f(\omega) $ is the Fourier transform of the time-modulation profile. 
Importantly, for infinitely large wavenumbers, $ q \rightarrow \infty $, Eq. (\ref{nonlocal-eps}) predicts that the material permittivity converges to that of free space regardless of temporal modulation. For instance, for the considered material (SiC with $\beta$ given in the caption of Fig. \ref{fig2p}), one can easily verify that $ \Delta \chi $ vanishes and $\epsilon_{bg}$ converges to unity for $ q \approx 10^8 \textrm{rad/m} \ll 2\pi/a_0 $, namely, the material response vanishes significantly before the relevant wavelength reaches the atomic scale, where the dipole approximation would break down.

Next, we examine the dispersion of surface phonon-polaritons (SPhPs) supported by the structure in Fig. \ref{fig1p}(a) in the presence and absence of nonlocality in the time-invariant case. The calculated local and nonlocal dispersion curves are shown in Fig. \ref{fig2p}. For small wavenumbers, Figs. \ref{fig2p}(a)-(b), the local and nonlocal cases are very similar, whereas significant differences appear at larger wavenumbers, Figs. \ref{fig2p}(c)-(d). In particular, in the local case the dispersion curves asymptotically converge to a flat line, which implies that (i) the material is still polarized for $ q \rightarrow \infty $, as discussed earlier, and (ii) it supports an infinite number of photonic states within a finite frequency range, leading to thermodynamic paradoxes and other issues, as extensively discussed in \cite{Optica,Fan,Monticone2020} in the plasmonic case. The inclusion of nonlocality in the polar dielectric material model resolves these issues by making the material response gradually vanish for very large wavenumbers and bending the dispersion asymptote downward, as seen in Fig. \ref{fig2p}(d). This downward bend originates from the red-shift of the phonon-polariton resonance frequency in the presence of nonlocality, i.e., $ \omega_0^2 \rightarrow \omega_0^2 - \beta^2q^2 $, which strongly affects the large-wavevector behavior of SPhPs. For large values of $q$ (until the dipole approximation holds), the dispersion is given by $ \omega = \sqrt{\omega_0^2- \beta^2q^2 + \omega_p^2/(1 + \epsilon_{\infty})} $, indicated by the dashed line in Fig. \ref{fig2p}(d). Interestingly, this behavior is qualitatively different from that of plasmonic systems where hydrodynamic nonlocality leads to a blue-shift, instead of a red-shift, in the plasma frequency and an upward bend of the plasmon-polariton dispersion asymptotes \cite{Polar-Dielectrics,Monticone2020,SM}.

\begin{figure}[t]
	\noindent \includegraphics[width=1.0\columnwidth]{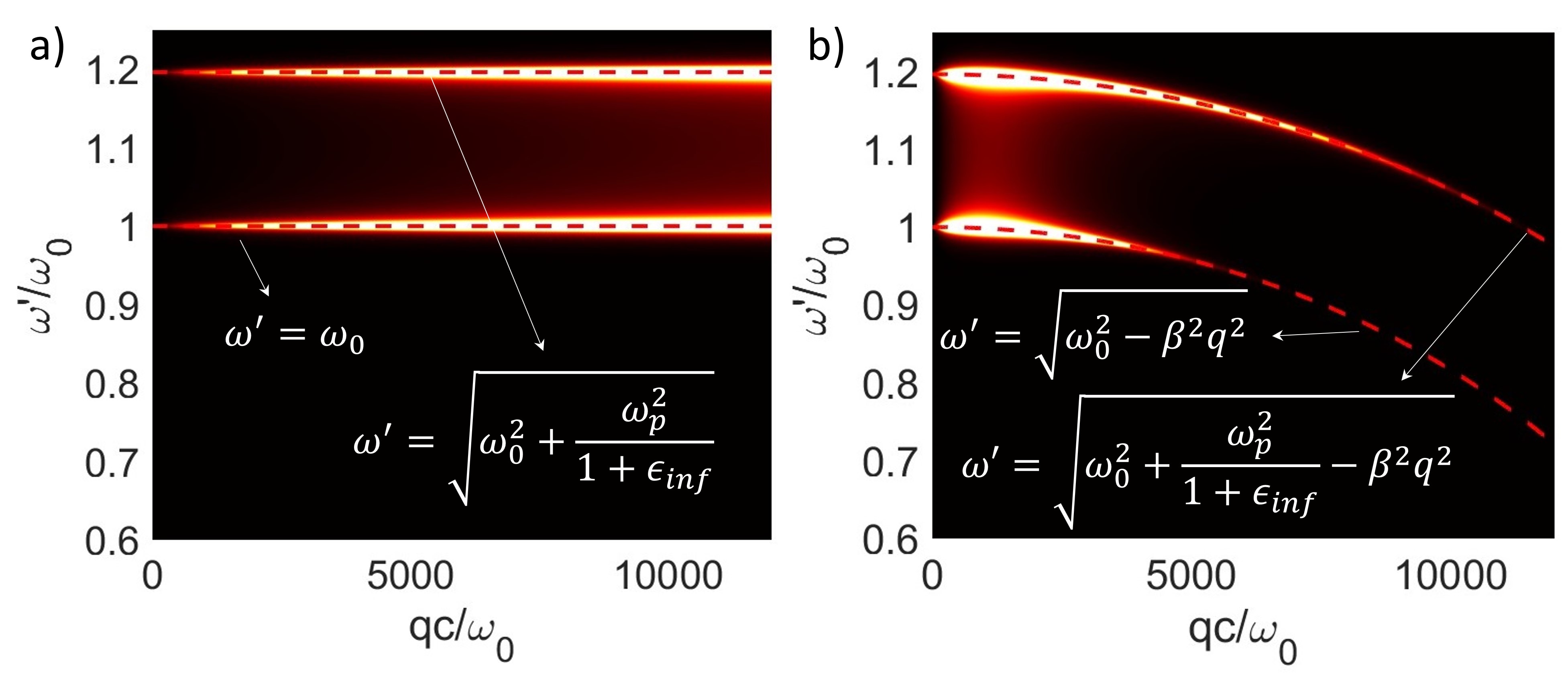}
	\caption{Wavevector- and frequency-resolved normalized emission rate (magnitude of integrand of Eqs. (\ref{P})), for (a) local ($ \beta = 0) $ and (b) nonlocal ($ \beta = 15.39\times 10^5~\mathrm{cm/s} $  \cite{Polar-Dielectrics}) cases.
	}
	\label{fig3p_int}
\end{figure}

These nonlocality-induced effects have several important consequences, both quantitative and qualitative, for the dynamical Casimir effects described above. To see this, we consider the case of SPhP pair generation when the top layer of a SiC slab (Fig. \ref{fig1p}(a)) is time-modulated with a resonant-frequency perturbation of the form $ f(t) =\cos (\Omega t) e^{-t^2/2T^2} $ where $T$ is the temporal width of the modulation, as in \cite{Jamison}. First, we show in Figs. \ref{fig3p_int}(a),(b) the wavevector- and frequency-resolved normalized emission rate (integrand of Eq. (\ref{P})), for the local and nonlocal cases, respectively. Different from the dispersion diagrams in Figs. \ref{fig2p}(c),(d), which simply indicate the location of the SPhP poles for the time-invariant slab, Figs. \ref{fig3p_int}(a),(b) indicate which parts of these dispersion diagrams contribute more strongly to the considered process, namely, the SPhP pair generation in the time-modulated case. Since this integrand is a function of three variables, the wavenumber $q$ and the pair frequencies $\omega$ and $\omega'$, in Figs. \ref{fig3p_int}(a),(b), we set $\omega= 1.2 \omega_0$ and plot the integrand as a density plot in terms of $\omega'$ and $q$ (for a short pulse modulation with $\Omega=2.2 \omega_0$ and $T=80 $ fs or $ T = 1.89 \left (2 \pi/\omega_0 \right ) $). We see from Fig. \ref{fig3p_int}(a) that, as expected, there are two main factors which cause a strong enhancement of the SPhP emission: (i) high photonic local density of states (LDOS) around the flat dispersion asymptote at $ \omega'/\omega_0 = 1.2 $, and (ii) the presence of a resonance in the material response around $ \omega'/\omega_0 = 1 $, where $|\Delta \chi|^2$ becomes large. This plot also clearly shows that, in the local case, the integral never converges as larger values of $q$ contributes more strongly to the pair emission rate. The nonlocal case (Fig. \ref{fig3p_int}(b)) is drastically different: both high-intensity features bend downward and the contribution of large wavevectors gradually vanish. Moreover, due to this downward bend, the range of frequencies $\omega'$ that strongly contribute to the emission process widens significantly for the same $\omega$ and modulation properties. 

\begin{figure}[b]
	\noindent \includegraphics[width=1.0\columnwidth]{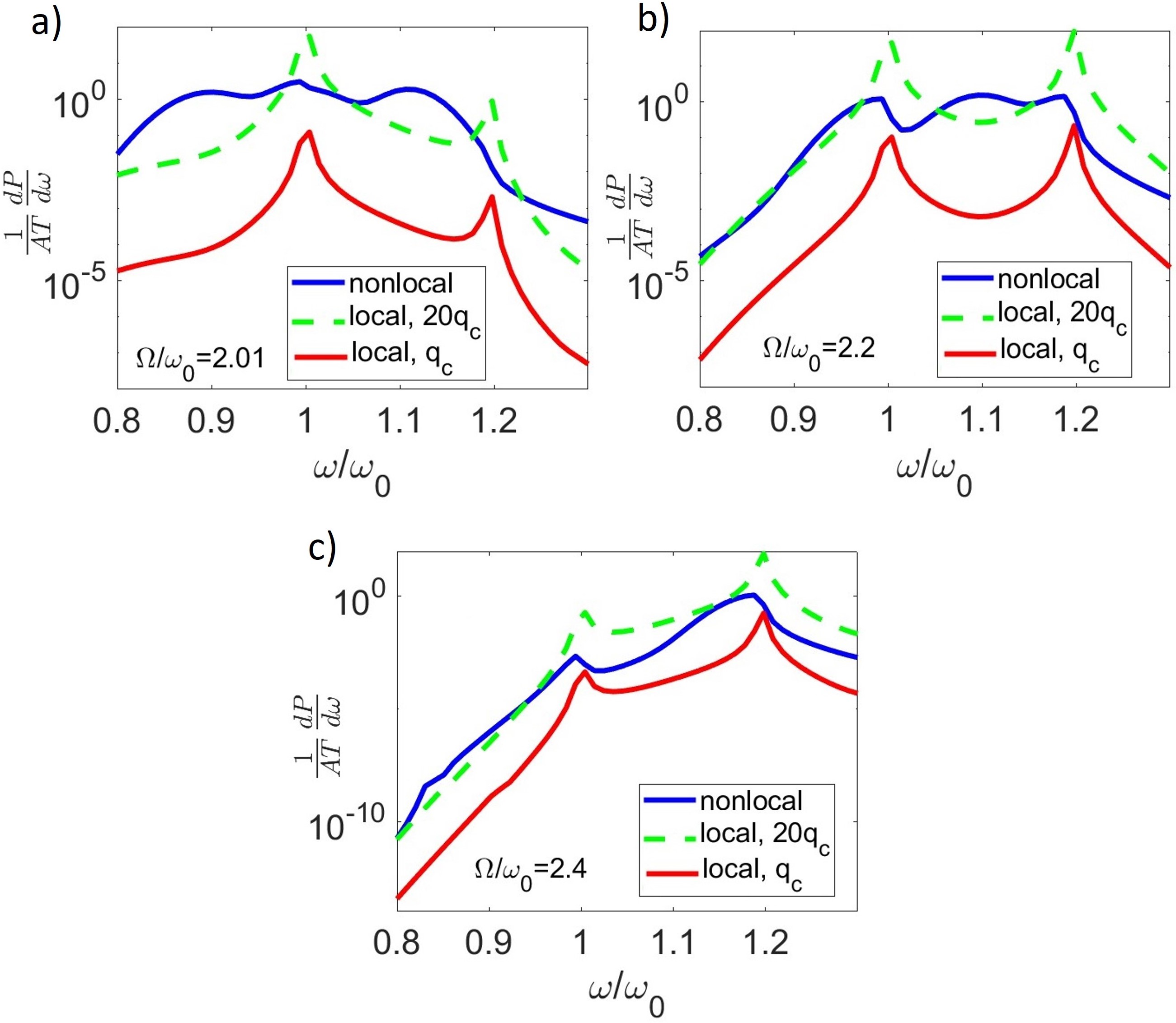}
	\caption{(a,b,c) Normalized emission rate distribution $ 1/(AT) dP/d\omega $, for the local and nonlocal cases, for different time-modulation frequencies: (a) $ \Omega/\omega_0 = 2.01 $, (b) $ \Omega/\omega_0 = 2.2 $, and (c) $ \Omega / \omega_0 = 2.4 $. The other temporal modulation parameters are as in Fig. \ref{fig3p_int}. 
	}
	\label{fig3p}
\end{figure} 

Finally, Figs. \ref{fig3p}(a), (b) and (c) show the normalized emission rate per frequency, $ 1/(AT) dP/d\omega $, for local and nonlocal SiC models for three different modulation frequencies . As discussed in \cite{Jamison}, this quantity corresponds to the emission rate which can be detected classically at frequency $ \omega $ (hence, without discriminating between the two emitted photons). 
In the local scenario, two different wavenumber cut-offs at $ q_c = 630\omega_0/c $ and $ 20 q_c$ were considered, truncating the integral in Eq. (\ref{P}) at these values. With these local models, the emission rate distribution is characterized by sharp peaks at $ \omega/\omega_0 = 1.2 $ and $ \omega/\omega_0 = 1$, as expected, and their relative strength can be modified by varying the central frequency $\Omega$ of the modulation pulse, as seen in Figs. \ref{fig3p}(a), (b) and (c). Under monochromatic time modulation, $ T \rightarrow \infty $, generated pair frequencies would respect the energy conservation constraint $ \omega + \omega' = \Omega $; therefore, as long as $ \Omega/2 $ lies between $ \omega/\omega_0 = 1.2 $ and $ \omega/\omega_0 = 1$, SPhP pair generation would take place at high rate, as both SPhPs can experience high LDOS and/or a strong material resonance. For a more realistic case with a modulation pulse with finite $ T $ (as considered here), and hence finite modulation spectrum, energy conservation can be satisfied over a wider range of frequencies $ \omega , \omega'$, but a similar behavior is expected as long as $T$ is not too small. As an example, by modulating the SiC material close to $\Omega/2 = \omega_0$, SPhP pairs are preferentially generated around $ \omega , \omega' \approx  \omega_0 $, corresponding to the main peak visible in Fig. \ref{fig3p}(c) (red curve), which physically originates from a parametric resonance of phonons \cite{Jamison}. Another peak around the frequency of the SPhP flat dispersion asymptote, $ \omega/ \omega_0 =1.2$, can also be observed, which is however orders of magnitude smaller. Conversely, for a modulation frequency close to $\Omega/2 = 1.2\omega_0$, SPhP pairs are preferentially generated around the asymptote frequency, as shown in Fig. 4(c). Finally, the two emission peaks can be equally enhanced (hence, with no preference for SPhP generation at these two frequencies) by tuning the modulation frequency in the middle of the range between material resonance and flat asymptote, $ \Omega/\omega_0 = 1.1 $, as in Fig. \ref{fig3p}(b).

While these results for the local case are sensible, they strongly depend on the choice of cut-off, which moderately affects their qualitative behavior and drastically change their numerical values, as seen by comparing the solid red and dashed green curves in Figs. \ref{fig3p}(a), (b) and (c). Considering nonlocal effects, as described above, solves this issue by providing a physically justified correction for the asymptotic behavior of time-modulated dispersive systems for large wavevectors. The solid blue lines in Fig. \ref{fig3p}(c), (d) and (e) show the corresponding emission rates for a model that includes nonlocality. The most evident qualitative difference is a significant broadening of the emission rate distribution --- the peaks are broader with significantly expanded tails --- which ultimately originates from the downward bend of the dispersion asymptote and of the material resonance condition (Figs. \ref{fig2p},\ref{fig3p_int}), and the resulting spreading of the LDOS. These physical effects are completely missed by the local model and suggest that strong emission rate can be obtained over a larger range of frequencies than the local model would suggest, but with less pronounced maxima, which are relevant features for future experimental observations.

In conclusion, our work shows that considering material nonlocality is necessary for an accurate, physically satisfactory and self-consistent description of dynamical Casimir effects in time-varying dispersive systems. 
More broadly, we believe that, as nonlocal effects have proved critical in understanding extreme effects in plasmonics \cite{Raza2015,Monticone2020,Khurgin,Optica,Fan}, they also represent an essential element in the study of dynamical quantum-vacuum effects in resonant nanophotonics.

\begin{acknowledgments}
	The authors acknowledge support from the Air Force Office of Scientific Research with Grant No. FA9550-22-1-0204, and the Office of Naval Research with Grant No. N00014-22-1-2486.
\end{acknowledgments}


\end{document}